\documentstyle[12pt]{article}
\begin{document}
\def\sss{\scriptscriptstyle}
\def\ss{\scriptstyle}
\def\endauthors{}
\def\authors#1\endauthors{#1}
\def\ket#1{|#1 \rangle}
\def\bra#1{\langle #1|}
\def\Ket#1{||#1 \rangle}
\def\Bra#1{\langle #1||}
\def\ov#1#2{\langle #1 | #2  \rangle }
\def\sixj#1#2#3#4#5#6{\left\{\negthinspace\begin{array}{ccc}
#1&#2&#3\\#4&#5&#6\end{array}\right\}}
\def\ninj#1#2#3#4#5#6#7#8#9{\left\{\negthinspace\begin{array}{ccc}
#1&#2&#3\\#4&#5&#6\\#7&#8&#9\end{array}\right\}}
\def\mbs{\mbox{\boldmath$\sigma$}}
\def\Ket#1{||#1 \rangle}
\def\Bra#1{\langle #1||}
\def\sss{\scriptscriptstyle}
\def\ss{\scriptstyle}
\def\rf#1{{(\ref{#1})}}
\def\mbs{\mbox{\boldmath$\sigma$}}
\def\a {{\alpha}}
\def\etal {{\it et al.}}
\begin{titlepage}
\pagestyle{empty}
\baselineskip=21pt
\vskip .2in
\begin{center}
{\large{\bf
Nuclear moments for the neutrinoless double beta decay}}
\end{center}

\vskip .1in
\authors
\centerline{C.~Barbero${}^{\dagger}$, F.~Krmpoti\'{c}${}^{\dagger}$}
\vskip .15in
\centerline{\it Departamento de F\'\i sica, Facultad de Ciencias}
\centerline{\it
Universidad Nacional de La Plata, C. C. 67, 1900 La Plata, Argentina.}
\vskip .1in
\centerline{and}
\centerline{D.~Tadi\'{c}}
\centerline{\it Physics Department, University of Zagreb}
\centerline{\it Bijeni\v cka c. 32-P.O.B. 162, 41000 Zagreb, Croatia.}
\endauthors
\endauthors
\vskip 0.5in
\centerline{ {\bf Abstract} }
\baselineskip=18pt
A derivation of the neutrinoless double beta decay rate, specially adapted
for the nuclear structure calculations, is presented. It is shown
that the Fourier-Bessel expansion of the hadronic currents, jointly with
the angular momentum recoupling, leads to very simple final expressions
for the nuclear form factors. This greatly facilitates the theoretical
estimate of the half life. Our approach does not require the closure
approximation, which however can be implemented if desired.
The method is exemplified for the $\beta\beta$ decay $^{48}Ca\rightarrow
\!^{48}Ti$, both within the QRPA and a shell-model like model.

\bigskip

\vspace{0.5in}
\noindent
$^{\dagger}$Fellow of the CONICET from Argentina.
\end{titlepage}
\baselineskip=18pt

\begin{center}
\section{Introduction}
\end{center}

The standard model (SM) of electroweak interactions has been brilliantly
confirmed by a host of experiments. But, about the
properties of neutrinos it says very little. Actually, in the SM it is
postulated that:
i) the neutrinos are the only fermions without the right-handed partners,
and
ii) their masslessness is dictated by the global lepton-number symmetry,
and not by a fundamental underlying principle, such as gauge invariance 
for the photon.
More, whether neutrinos behave so "trivially" as required by the SM
is one of the most fundamental open questions of the present-day
physics.

It has been known for a long time \cite{Hax84,Doi85,Ver86,Tom86,Doi93}
that neutrinoless double beta decay ($\beta\beta_{0\nu}$) is a very
sensitive probe of lepton number violating terms in the Lagrangian such as
the Majorana mass of the light neutrinos, right--handed weak couplings as
well as the Higgs exchange \cite {Moh81},
right-handed weak coupling involving heavy Majorana neutrinos \cite{Hal76},
massless Majoron emission \cite{Doi85,Gel81,Bur93,Bam95,Bar97},
and R--parity breaking in the supersymmetric model
\cite{Ver87,Hir95}.
Thus, if the $\beta\beta_{0\nu}$ decay is someday observed experimentally
it would hint new physics beyond the SM.
But, even if it is not observed, the measured limits on its
transition probability, which are steadily improving \cite{Kla94}, could
be translated into more stringent constraints on the parameters of the
just mentioned new theoretical developments.

Yet, the extraction of these constraints from the data is only possible
when we know how to deal with the nuclear structure involved in the
$\beta\beta_{0\nu}$ decay. This is not  at all an easy task, because of:

(1) the nuclear hamiltonian is only roughly known to the extent that
the choice of the appropriate parametrization is an  art,

(2) there is in general a very large amount of nuclear states involved in
the calculation, and

(3) the formulas for the $\beta\beta_{0\nu}$ decay rate are rather complex
and difficult to implement in a nuclear structure calculation.

In the present work we derive simple expressions for the nuclear matrix
elements, especially tailored for the nuclear structure calculations.
The simplification mainly comes from the Fourier-Bessel expansion
of the term $\exp[i{\bf k}\cdot({\bf r}_1-{\bf r}_2)]$
in the transition amplitude, and in
doing the integrations in the following order: first on $d\Omega_{{\bf k}}$,
then on $d{\bf r}_1$ and $d{\bf r}_2$, and finally on $k^2dk$ \cite{Hor61}.
So far, the same procedure has been applied for the evaluation of the
matrix elements $M_{{\sss {F}}}$,  $M_{{\sss {GT}}}$ \cite{Krm92,Krm94} 
and $M_{{\sss {R}}}$ \cite{Bar97} that arise from the electron $s$-wave. 
Here we also dealt with the $p$-wave matrix elements that
are relevant when the admixture of the right-hand lepton current is 
considered.
Other studies on the subject are those of Vergados {\it et al.}
\cite{Ver90}, who derived the formulas directly in the momentum space, and
those of Suhonen, Khadkikar and Faessler \cite{Suh91}, who worked in a
framework of a relativistic quark confinement model.

This paper is organized as follows: In sec. 2 we discuss the basic
mechanism for the $\beta\beta_{0\nu}$ decay,
presenting  the effective
hamiltonian and the transition amplitude in the form convenient
for the multipole expansion, which is carried out in sec. 3.
In sec. 4 we give the detailed formulas for the nuclear matrix elements
and discuss the nuclear structure calculations involved in the problem.
Summarizing conclusions are drawn in sec. 5.

\begin{center}
\section{Effective Hamiltonian and the half life}
\end{center}

The $0\nu\beta\beta$ half life
\begin{equation}
[T_{0\nu}(0^+\rightarrow  0^+)]^{-1}=\frac{\Gamma_{0\nu}}{\ln 2},
\label{1}\end{equation}
for the decay from the state ${\ket{{\ss {I}}}}$
in the $(N,Z)$ nucleus to the state ${\ket{{\ss {F}}}}$ in the 
$(N-2,Z+2)$ nucleus (with energies $E_I$ and $E_F$ and spins and 
parities $J^{\pi}=0^+$), is evaluated via the second order Fermi's 
golden rule. Thus the decay rate (in $\hbar=c=m_e$ units) is \cite{Pri59}
\begin{equation}
\Gamma_{0\nu}=2\pi\sum_{s_{e_1}s_{e_2}}\int|R_{0\nu}(e_1,e_2)|^2
\delta(\epsilon_1+\epsilon_2+E_{\sss F} -E_{\sss I})
\frac{d{\bf p}_1}{(2\pi)^3}\frac{d{\bf p}_2}{(2\pi)^3},
\label{2}\end{equation}
with
\begin{equation}
R_{0\nu}(e_1,e_2)=\sum_{\sss N}\sum_{s_\nu}\int\frac{d{\bf k}}{(2\pi)^3}
\frac{\bra{{\sss F};e_1,e_2}H_{\sss W}\ket{{\ss N};e_1,\nu}
\bra{{\ss N};e_1,\nu}H_{\sss W}\ket{{\ss I}}}
{E_{\sss I}-E_{\sss N}-\epsilon_1-{\omega}},
\label{3}\end{equation}
where $e\equiv (\epsilon,{\bf p},s_e)$ 
($\nu\equiv ({\omega},{\bf k},s_\nu)$) stands for the
energy, momentum and spin projection of the electron (neutrino), and
${\ss N}$ runs over all levels in the $(N-1,Z+1)$ nucleus.

The effective weak hamiltonian reads \cite{Doi85,Tom86,Doi93}
\begin{equation}
H_{\sss W}=\frac{G}{\sqrt{2}}\sum_{\ell=1}^{2n}\int d{\bf x}
[j_{L\ell\mu}({\bf x}) {\tilde{J}}_{L\ell}^{\mu{\dagger}}({\bf x})
+j_{R\ell\mu}({\bf x}){\tilde{J}}_{R\ell}^{\mu{\dagger}}({\bf x})+h.c.],
\label{4}\end{equation}
where the summation goes on the number on lepton generations,
\begin{eqnarray}
j_{L,R\ell}^{\mu}({\bf x})&=&2\bar{\Psi}({\bf x}){\gamma}^\mu 
P_{L,R}{\sf N}_{L,R\ell}({\bf x});
\hspace{1cm} P_{L,R}=\frac{1}{2}(1\mp{\gamma}_5),
\label{5}\end{eqnarray}
are the leptonic currents, formed out from the electron field
$\Psi({\bf x})$ and the Majorana neutrino field ${\sf N}_\ell({\bf x})$ 
of mass $m_\ell$, and 
\begin{equation}
{\tilde{J}}_{L\ell}^{\mu{\dagger}}({\bf x})=U_{e\ell}
J_L^{\mu{\dagger}}({\bf x}),\hspace{1cm}
{\tilde{J}}_{R\ell}^{\mu{\dagger}}({\bf x})
=V_{e\ell}(\lambda J_R^{\mu{\dagger}}({\bf x})+
\eta J_L^{\mu{\dagger}}({\bf x})),
\label{6}\end{equation}
contain the hadronic $(V\mp A)$ currents $J_{L,R}^{\mu}$.
\footnote{We do not consider the admixture of the hadronic $(V+A)$
current into ${\tilde{J}}_{L}^{\mu}$, since its contribution to
the $\beta\beta$ decay amplitudes is negligible \cite{Tom86}.}
$U_{e\ell}$ and $V_{e\ell}$ are the neutrino mixing matrices for the left-
and right-handed sectors, and $\lambda$
and $\eta$ are the strengths of admixtures of the $(V+A)$ current.

Within the non-relativistic impulse approximation the hadronic currents
read,
\begin{eqnarray}
J_{L,R}^\mu({\bf x})&=&\left(\frac{}{}\rho_{{\sss {V}}}({\bf x})\mp
\rho_{{\sss {A}}}({\bf x}),
{\bf j}_{{\sss {V}}}({\bf x})\mp{\bf j}_{{\sss {A}}}({\bf x})\right)
\label{7}\end{eqnarray}

where
\footnote{See eq. (3D-18) in ref. \cite{Boh69}. The correspondence between
the non-relativistic approximations
used here and that prevailingly employed in the studies of the
$\beta\beta_{0\nu}$ decay
\cite{Hax84,Doi85,Ver86,Tom86,Doi93},
can be find on p. 516 of the Walecka's book \cite{Wal95}.}
\begin{eqnarray}
\rho_{{\sss {V}}}({\bf x})&=&g_{{\sss {V}}} \sum_n\tau_n^+ 
\delta({\bf x}-{\bf r}_n),\nonumber\\
\rho_{{\sss {A}}}({\bf x})&=&\frac{g_{{\sss {A}}}}{2M_{\sss N}}
\sum_n\tau_n^+[\mbs_n\cdot{\bf p}_n\delta({\bf x}-{\bf r}_n)
+\delta({\bf x}-{\bf r}_n) \mbs_n\cdot{\bf p}_n] ,\label{8}\\
{\bf j}_{{\sss {V}}}({\bf x})&=&\frac{g_{{\sss {V}}}}{2M_{\sss N}}
\sum_n\tau_n^+[{\bf p}_n\delta({\bf x}-{\bf r}_n)
+\delta({\bf x}-{\bf r}_n){\bf p}_n+f_{\sss W}
{\mbox{\boldmath$\nabla$}}{\times}\mbs_n
\delta({\bf x}-{\bf r}_n)]  ,\nonumber\\
{\bf j}_{{\sss {A}}}({\bf x})&=&g_{{\sss {A}}}\sum_n\tau_n^+ 
\mbs_n\delta({\bf x}-{\bf r}_n),
\nonumber\end{eqnarray}
are the one-body vector ($V$) and axial-vector ($A$) densities and
currents, $M_{\sss N}$ is nucleon mass and $f_{\sss W}=4.7$ is the
effective weak-magnetism coupling constant.

Merging \rf{4} into \rf{3} and performing the $s_\nu$-summation one
gets \cite{Tom86}:
\begin{eqnarray}
R_{0\nu}&=&\frac{G^2}{\sqrt{2}}
\sum_{\ell=1}^{2n}\sum_{\sss N}\sum_{\a,\beta=L,R}\int d{\bf x} d{\bf y} \int
\frac{d{\bf k}}{(2\pi)^3}
\bra{{\ss{F}}}{\tilde{J}}_{\beta\ell}^{\nu{\dagger}}({\bf y})
e^{i{\bf k}\cdot{\bf y}}\ket{{{\ss{N}}}} \bra{{{\ss{N}}}}
{\tilde{J}}_{\a \ell}^{\mu{\dagger}}({\bf x})e^{-i{\bf k}\cdot{\bf x}}
\ket{{\ss{I}}}\nonumber\\
&{\times}&[1-P(e_1,e_2)]\frac{\bar{\psi}(\epsilon_2,{\bf y}){\gamma}_\nu 
P_\beta\left({\omega}{\gamma}^0-{\bf k}\cdot{\mbox{\boldmath$\gamma$}}
+m_\ell\right)P_\a{\gamma}_\mu{\psi}^C(\epsilon_1,{\bf x})}
{{\omega}(\epsilon_1+{\omega}+E_{{\sss N}}-E_{{\sss I}})},
\label{9}\end{eqnarray}
where $\psi(\epsilon_1,{\bf x})$ and
$\psi(\epsilon_2,{\bf x})$ are the wave functions of the emitted
electrons, and the operator $P(e_1,e_2)$ interchanges the particles $e_1$ 
and $e_2$. The structure of the eq. \rf{9} suggests that it might be 
convenient to introduce the Fourier transforms of the quantities defined 
in \rf{8}, {\em i.e., }
\begin{eqnarray}
\rho({\bf k})&=&\int d{\bf x}\rho({\bf x})
e^{-i{\bf k}\cdot{\bf x}},\nonumber\\
{\bf j}({\bf k})&=&\int d{\bf x}{\bf j}({\bf x})
e^{-i{\bf k}\cdot{\bf x}}.
\label{10}\end{eqnarray}

Next, ensuing the usual procedure \cite{Doi85,Tom86,Doi93}, we evaluate
the $s_{1/2}$ and $p_{1/2}$ contributions of the electron wave functions
to the amplitude $R_{0\nu}$. The first ones give rise to the following 
${\bf k}$ and ${\ss N}$ dependent nuclear moments
\begin{eqnarray}
{\sf M}_{{\sss F}}({\bf k},{{\ss N}})&=&
{\bra{{\ss F}}}\rho_{{\sss {V}}}(-{\bf k})\ket{{\ss N}}\bra{{\ss N}}
\rho_{{\sss {V}}}({\bf k}){\ket{{\ss I}}},\nonumber\\
{\sf M}_{{\sss GT}}({\bf k},{{\ss N}})&=&
{\bra{{\ss F}}}{\bf j}_{{\sss {A}}}(-{\bf k})\ket{{\ss N}}\cdot
\bra{{\ss N}}{\bf j}_{{\sss {A}}}({\bf k}){\ket{{\ss I}}},\label{11}\\
{\sf M}_{{\sss R}}({\bf k},{{\ss N}})&=&-i{\sf R}{\bf k}\cdot{\bra{{\ss F}}}
{\bf j}_{{\sss {A}}}(-{\bf k})
\ket{{\ss N}}{\times}\bra{{\ss N}}{\bf j}_{{\sss {V}}}({\bf k})
{\ket{{\ss I}}},
\nonumber\end{eqnarray}
where ${\sf R}$ is the nuclear radius, and the second one to
\begin{eqnarray}
{\sf M}_{{\sss F'}}({\bf k},{{\ss N}})&=&2\sqrt{3}i{\bra{{\ss F}}}
\rho^{(0)}_{{\sss{V}}}(-{\bf k})
\ket{{\ss N}}\bra{{\ss N}}\rho_{{\sss {V}}}({\bf k}){\ket{{\ss I}}},
\nonumber\\
{\sf M}_{{\sss GT'}}({\bf k},{{\ss N}})&=&6i
{\bra{{\ss F}}}j^{(01)}_{{\sss{A}}}(-{\bf k})\ket{{\ss N}}\cdot
\bra{{\ss N}}{\bf j}_{{\sss {A}}}({\bf k}){\ket{{\ss I}}},\label{12}\\
{\sf M}_{{{\sss T}}}({\bf k},{{\ss N}})&=&2i
{\bra{{\ss F}}}j^{(21)}_{{\sss{A}}}(-{\bf k})\ket{{\ss N}}\cdot
\bra{{\ss N}}{\bf j}_{{\sss {A}}}({\bf k}){\ket{{\ss I}}},\nonumber\\
{\sf M}_{{{\sss P}}}({\bf k},{{\ss N}})&=&-\sqrt{2}i[{\bra{{\ss F}}}
j^{(10)}_{{\sss{A}}}(-{\bf k})
\ket{{\ss N}}\bra{{\ss N}}\rho_{{\sss {V}}}({\bf k}){\ket{{\ss I}}}
-{\bra{{\ss F}}}{\bf j}_{{\sss {A}}}(-{\bf k})\ket{{\ss N}}\cdot
\bra{{\ss N}}\rho_{{\sss{V}}}^{(1)}({\bf k}){\ket{{\ss I}}}],
\nonumber\end{eqnarray}
where we have introduced the tensor operators
\begin{eqnarray}
\rho^{(J)}({\bf k})&=&\int d{\bf x}\rho({\bf x})
({\bf k}\otimes{\bf x})^{(J)} e^{-i{\bf k}\cdot{\bf x}},\nonumber\\
\nonumber\\j^{(LJ)}({\bf k})&=&\hat{L}\hat{J}^{-1}
\int d{\bf x}[{\bf j}({\bf x})\otimes
({\bf k}\otimes{\bf x})^{(L)}]^{(J)}e^{-i{\bf k}\cdot{\bf x}},
\label{13}\end{eqnarray}
with $\hat{L}=\sqrt{2L+1}$. The explicit form of the matrix elements defined
in \rf{11} and \rf{12} are shown in the appendix A.

We now define the nuclear matrix elements
\begin{eqnarray}
M_{{{\sss X}}}&=&\frac{{\sf R}}{4\pi g_{{\sss {A}}}^2}\sum_{\sss N}\int d{\bf k}
v(k,\omega_{{\sss {N}}})
{\sf M}_{{\sss X}}({\bf k},{{\ss N}})~~~~\mbox{for}~~~~{{\ss X}}={{\ss F}},
{{\ss GT}},{{\ss F'}},{{\ss GT'}},
{{\ss P}},{{\ss R}},{{\ss T}},
\label{14}\end{eqnarray}
and
\begin{eqnarray}
M_{{{\sss X}{\omega}}}&=&\frac{{\sf R}}{4\pi g_{{\sss {A}}}^2}\sum_{\sss N}\int d{\bf k}
v_{{\omega}}(k,\omega_{{\sss {N}}})
{\sf M}_{{\sss X}}({\bf k},{{\ss N}})
~~~~\mbox{for}~~~~{{\ss X}}={{\ss F}},{{\ss GT}},
\label{15}\end{eqnarray}
with
\begin{equation}
v(k,\omega_{{\sss {N}}})=\frac{2}{\pi}
\frac{1}{k(k+\omega_{{\sss {N}}})},~~~~~
v_{{\omega}}(k,\omega_{{\sss {N}}})=\frac{2}{\pi}
\frac{1}{(k+\omega_{{\sss {N}}})^2},
\label{16}\end{equation}
and
\begin{equation}
\omega_{{\sss {N}}}=E_{{\sss N}}-\frac{1}{2}\left(E_{{\sss I}}
+E_{{\sss F}}\right).
\label{17}\end{equation}
In deriving the expression \rf{17} we have approximated the electron
energies as $\epsilon_{1,2}\cong (E_{{\sss I}}-E_{{\sss F}})/2$. We 
have also neglected the neutrino mass in comparison with $k$, 
{\em i.e., } we have taken ${\omega}\cong k$.

For the transition amplitude we get
\begin{equation}
R_{0\nu}(e_1,e_2)=\frac{g_{{\sss {A}}}^2G^2}
{4\pi {\sf R}\sqrt{2}}\sum_{k=1}^5Z_k L_k(\epsilon_1,\epsilon_2),
\label{18}\end{equation}
where
\begin{eqnarray}
Z_1&=&<m_\nu>(M_F-M_{GT}),\nonumber\\
Z_2&=&<\eta>(M_{GT{\omega}}+M_{F{\omega}})+<\lambda>(M_{F{\omega}}
-M_{GT{\omega}}),\nonumber\\
Z_3&=&4<\eta>M_R,\label{19}\\
Z_4&=&\frac{2}{3}i[<\lambda>({M'}_{GT}-6M_{T}+3{M'}_{F})
-<\eta>({M'}_{GT}-6M_{T}-3{M'}_{F})],\nonumber\\
Z_5&=&4i<\eta>M_{P},
\nonumber\end{eqnarray}
encompass the hadronic matrix elements, as well as the parameters
\begin{eqnarray}
<m_\nu>&=&{\sum_\ell}' m_\ell U_{e\ell}^2,\nonumber\\
<\lambda>&=&\lambda{\sum_\ell}' U_{e\ell}V_{e\ell},\label{20}\\
<\eta>&=&\eta{\sum_\ell}' U_{e\ell}V_{e\ell},
\nonumber\end{eqnarray}
where the summation
${\sum_\ell}'$ goes only on the light neutrinos
\cite{Tom86,Doi93}.
The leptonic matrix elements $L_1(\epsilon_1,\epsilon_2)$ are
displayed in the appendix B.

Finally, by performing the integrations (summations) on the electron states
indicated in \rf{2}, we obtain the familiar expression for
the $0\nu\beta\beta$ half life \cite{Doi85,Tom86}
\begin{eqnarray}
[T_{0\nu}(0^+\rightarrow  0^+)]^{-1}
&=&<m_\nu>^2C_1+<\lambda>^2C_2+<\eta>^2C_3\\
&+&<m_\nu><\lambda>C_4+<m_\nu><\eta>C_5+<\lambda><\eta>C_6,
\nonumber
\label{21}\end{eqnarray}
where
\begin{eqnarray}
C_1&=&(M_F-M_{GT})^2{\cal G}_{1},\nonumber\\
C_2&=&M_{2-}^2{\cal G}_2+\frac{1}{9}M_{1+}^2{\cal G}_4
-\frac{2}{9}M_{2-}M_{1+}{\cal G}_3,\nonumber\\
C_3&=&M_{2+}^2{\cal G}_2+\frac{1}{9}M_{1-}^2{\cal G}_4
-\frac{2}{9}M_{2+}M_{1-}{\cal G}_3+M_R^2{\cal G}_9+M_RM_{P}{\cal G}_7
+M_{P}^2{\cal G}_8,\nonumber\\
C_4&=&(M_F-M_{GT})\left[M_{2-}{\cal G}_3-M_{1+}{\cal G}_4\right],
\label{22}\\
C_5&=&-(M_F-M_{GT})\left[M_{2+}{\cal G}_3-M_{1-}{\cal G}_4
+M_R{\cal G}_6+M_{P}{\cal G}_5\right],\nonumber\\
C_6&=&-2M_{2-}M_{2+}{\cal G}_2+\frac{2}{9}\left[M_{2-}M_{1-}+
M_{2+}M_{1+}\right]{\cal G}_3-\frac{2}{9}M_{1-}M_{1+}{\cal G}_4,
\nonumber\end{eqnarray}
contain the usual combinations of the matrix elements
\begin{eqnarray}
M_{1\pm}&=&{M'}_{GT}-6M_{T}\pm 3{M'}_{F},\nonumber\\
M_{2\pm}&=&M_{GT{\omega}}\pm M_{F{\omega}}-\frac{1}{9}M_{1\mp},
\label{23}\end{eqnarray}
and the kinematical factors
\begin{equation}
{\cal G}_k=\frac{g_{{\sss {A}}}^4G^4}{32{\sf R}^2\pi^5\ln~2}
\left(\frac{2\pi\a Z}{1-e^{-2\pi\a Z}}\right)^2{\cal F}_k(T_0).
\label{24}\end{equation}
The electron phase-space factors ${\cal F}_k(T_0)$,
as a function of the maximum kinetic energy $T_0=E_{\sss I}-E_{\sss F}-2$,
are listed in the appendix B.

It might be important to stress that, within the procedure followed 
here to derive the result \rf{18}, we do not need to recur at all to
so called closure approximation (CA). (Remind that the
CA implies:
i) to supplant the energies $E_{\sss N}$ by an average values
$<E_{{\sss N}}>$, and
ii) to use the closure relation $\sum_{{\sss N}}\ket{{{\ss N}}}
\bra{{{\ss N}}}=1$ for the intermediate states.)
When reworked in the CA, the moments \rf{14} and \rf{15} are directly
comparable with those that appear in the literature
\cite{Hax84,Doi85,Ver86,Tom86,Doi93}.

\begin{center}
\section{Multipole expansion and angular momentum recoupling}
\end{center}

        The starting point for the multipole expansion of the hadronic
current is to use
the Fourier-Bessel relation
\begin{eqnarray}
e^{i{\bf k}\cdot{\bf r}}
&=&4\pi\sum_{L}i^{L}j_L(kr)(Y_{L}(\hat{\bf k})\cdot Y_{L}(\hat{\bf r}))
\nonumber\\
&\equiv& 4\pi\sum_{L}i^{L}(-1)^L\hat{L}j_L(kr)
[Y_{L}(\hat{\bf k})\otimes Y_{L}(\hat{\bf r})]_0,
\label{25}\end {eqnarray}
in the equations exhibited in the appendix A. Then we perform the angular momentum recoupling,
and rewrite the nuclear moments \rf{11} and \rf{12} in terms of the
one-body spherical tensor operators
\begin{eqnarray}
{\sf Y}_{\lambda JM}^{\kappa}(k)&=&\sum_n\tau_n^+r_n^\kappa
j_\lambda(kr_n)Y_{JM}(\hat{\bf r}_n),\nonumber\\
{\sf S}_{\lambda LJM}^{\kappa}(k)&=&\sum_n
\tau_n^+r_n^{\kappa}j_\lambda(kr_n)[\mbs_n\otimes 
Y_{L}(\hat{\bf r}_n)]_{JM},\nonumber\\
{\sf P}_{LJM}(k)&=&\sum_n\tau_n^+j_L(kr_n)[{\bf p}_n\otimes
Y_{L}(\hat{\bf r}_n)]_{JM}.
\label{26}\end{eqnarray}
Finally, the angular integration on $d\Omega_{\bf k}$ is done.
We illustrate the procedure by sketching
in the appendix C,
the derivation of a part
of the final formula for the nuclear matrix element $M_{{\sss R}}$.
Proceeding in a similar way with the remaining matrix elements we obtain:
\begin{eqnarray}
M_{{\sss F}}&=&4\pi{\sf R}\left(\frac{g_{{\sss V}}}{g_{{\sss A}}}\right)^2
\sum_{J{\sss N}} \int v(k,\omega_{{\sss {N}}})k^2dk {\bra{{\ss F}}}
{\sf Y}_{JJ}^0(k)
{\ket{{\ss N}}}\cdot{\bra{{\ss N}}}{\sf Y}_{JJ}^0(k){\ket{{\ss I}}},
\label{27}\\
M_{{\sss GT}}&=&4\pi{\sf R} \sum_{LJ{\sss N}}(-1)^{1+L+J}\int
v(k,\omega_{{\sss {N}}})k^2dk
{\bra{{\ss F}}}{\sf S}_{LLJ}^0(k){\ket{{\ss N}}}\cdot 
{\bra{{\ss N}}}{\sf S}_{LLJ}^0(k){\ket{{\ss I}}},
\label{28}\\
M_{{\sss F'}}&=&-8\pi{\sf R}\left(\frac{g_{{\sss V}}}{g_{{\sss A}}}\right)^2
\sum_{LJ{\sss N}}i^{L-J+1}(J1|L)(J1|L)\nonumber\\
&{\times}&\int v(k,\omega_{{\sss {N}}})k^3dk
\bra{{\ss F}}{\sf Y}_{LJ}^1(k){\ket{{\ss N}}}\cdot
{\bra{{\ss N}}}{\sf Y}_{JJ}^0(k){\ket{{\ss I}}},
\label{29}\\
M_{{\sss GT'}}&=&8\pi{\sf R}\sum_{LL'J{\sss N}}i^{L-L'+1}(-1)^{L'+J}(L'1|L)(L'1|L)
\nonumber\\
&{\times}&\int v(k,\omega_{{\sss {N}}})k^3dk
{\bra{{\ss F}}}{\sf S}_{LL'J}^1(k){\ket{{\ss N}}}\cdot{\bra{{\ss N}}}
{\sf S}_{L'L'J}^0(k){\ket{{\ss F}}},
\label{30} \end{eqnarray}
\begin{eqnarray}
M_{{\sss R}}&=&\frac{2\pi{\sf R}^2}{M_{\sss N}} \frac{g_{{\sss V}}}{g_{{\sss A}}}
\sum_{LL'J{\sss N}}i^{L+L'}(-1)^J\int v(k,\omega_{{\sss {N}}})k^3dk
\bra{{\ss F}}{\sf S}_{LLJ}^0(k)\ket{{\ss N}}\cdot\nonumber\\
&&\left\{f_{\sss W} k\left[\frac{}{}\delta_{LL'}-(J1|L)(J1|L')\right]
\bra{{\ss N}} {\sf S}_{L'L'J}^0(k)\ket{{\ss I}}\right.\nonumber\\
&-&\left.2\sqrt{6}(-1)^{L+J}\hat{L}\sixj{L}{J}{1}{1}{1}{L'}(L1|L')
\bra{{\ss N}}{\sf P}_{L'J}(k)\ket{{\ss I}}\right\},
\label{31} \end{eqnarray}
\begin{eqnarray}
M_{{\sss T}}&=&40\pi{\sf R}\sum_{LL'J'J{\sss N}}i^{L+L'+1}{\hat{L}^2}
(1L|J')(1L|L')
\sixj{1}{2}{1}{J'}{L}{L'}\sixj{1}{2}{1}{J'}{J}{L'}\nonumber\\
&{\times}& \int v(k,\omega_{{\sss {N}}})k^3dk {\bra{{\ss F}}}
{\sf S}_{LJ'J}^1(k){\ket{{\ss N}}}\cdot{\bra{{\ss N}}}
{\sf S}_{L'L'J}^0(k){\ket{{\ss I}}},
\label{32}\\
M_{{\sss P}} &=&8\pi\sqrt{6}{\sf R}\frac{g_{{\sss A}}}{g_{{\sss V}}}
\sum_{LJ{\sss N}}i^{L+J+1}\hat{J}(J1|L)(J1|L)\sixj{J}{L}{1}{1}{1}{J}\nonumber\\
&{\times}&\int v(k,\omega_{{\sss {N}}})k^3dk{\bra{{\ss F}}}
{\sf S}_{LJJ}^1(k){\ket{{\ss N}}}\cdot{\bra{{\ss N}}} 
{\sf Y}_{JJ}^0(k){\ket{{\ss I}}},
\label{33}\end{eqnarray}
where $(L1|J)$ is a short notation for the Clebsh-Gordon coefficient
$(L010|J0)$. The formulas for the matrix elements $M_{{\sss F}{\omega}}$
and  $M_{{\sss GT}{\omega}}$ are obtained from those for $M_{{\sss F}}$ and
$M_{{\sss GT}}$ with the replacement $v(k,\omega_{{\sss {N}}})\rightarrow
v_{{\omega}}(k,\omega_{{\sss {N}}})$  (see eqs. \rf{14} and \rf{15}).

The evaluation of the
$0\nu\beta\beta$ matrix elements encompasses:

i) the appraisal of the scalar product
\begin{equation}
{\bra{{\ss F}}}{\sf T}_{J}(k){\ket{{\ss N}}}\cdot
{\bra{{\ss N}}}{\sf T}_{J}(k){\ket{{\ss I}}},
\label{34}\end{equation}
where ${\sf T}_{J}(k)$ represents any of the one-body operators displayed
in \rf{26}, and

ii) the integration on the neutrino momentum $k$.

More details on these two steps are given in the next section.
\newpage
\begin{center}
\section{Nuclear structure calculations}
\end{center}

To evaluate the matrix elements \rf{34} it is convenient to rewrite the
operators \rf{26} from the Hilbert space to the Fock space \cite{Boh69},
{\em i.e.},
\begin{equation}
{\sf T}_{JM}(k)=\hat{J}^{-1}\sum_{pn}\Bra{p} {\sf T}_{J}(k)\Ket{n}
\left(a^{{\dagger}}_p a_{\bar{n}}\right)_{JM}.
\label{35}\end{equation}
In this way we get
\begin{eqnarray}
&&\sum_M{\bra{{\ss F}}}{\sf T}_{J}(k){\ket{{\ss N}}}\cdot
{\bra{{\ss N}}}{\sf T}_{J}(k){\ket{{\ss I}}}
\equiv\sum_{\a\pi M}{\bra{{0^+_f}}}{\sf T}_{J}(k){\ket{J^\pi_{\a} M}}\cdot
{\bra{J^\pi_{\a} M}}{\sf T}_{J}(k){\ket{0^+_i}}\nonumber\\
&=&(-)^J\sum_{\a\pi pnp'n'}
\Bra{p} {\sf T}_{J}(k)\Ket{n}\rho^{ph}(pnp'n';J_\a^\pi)
\Bra{p'} {\sf T}_{J}(k)\Ket{n'},
\label{36}\end{eqnarray}
where
\begin{eqnarray}
\rho^{ph}(pnp'n';J_\a^\pi)
&=&\hat{J}^{-2}
\Bra{0^+_f} (a^{{\dagger}}_{p}a_{\bar{n}})_{J^\pi}
\Ket{{J_\a^\pi}}
\Bra{{J_{\a}^\pi}}(a^{{\dagger}}_{p'} a_{\bar{n}'})_{J^\pi}
\Ket{ 0^+_i},
\label{37}\end{eqnarray}
is a two-body state dependent particle-hole (ph) density matrix, and
the index $\a$ labels different intermediate states with
the same spin $J$ and parity $\pi$.

Within the CA we can sum over $\a$, and deal with the
state independent ph density matrix
\begin{eqnarray}
\rho^{ph}_{cl}(pnp'n';J^\pi) =\sum_{\a}\rho^{ph}(pnp'n';J_\a^\pi)
\equiv \hat{J}^{-1}
\bra{0^+_f}\left[(a^{{\dagger}}_{p}a_{\bar{n}})_{J^\pi}
(a^{{\dagger}}_{p'} a_{\bar{n}'})_{J^\pi}\right]_0\ket{ 0^+_i},
\label{38}\end{eqnarray}
which is related with the particle-particle (pp) density matrix
\begin{eqnarray}
\rho^{pp}(pp'nn';J^\pi) = \hat{J}^{-1}
\bra{0^+_f} \left[(a^{{\dagger}}_{p}a^{{\dagger}}_{p'})_{J^\pi}
(a_{\bar{n}} a_{\bar{n}'})_{J^\pi}\right]_0\ket{0^+_i},
\label{39}\end{eqnarray}
by a Pandya like relation
\begin{eqnarray}
\rho^{ph}_{cl}(pnp'n';J^\pi) =\sum_{I^\pi}(-)^{j_n+j_{p'}+J+I}
\hat{I}^2\sixj{j_p}{j_n}{J}{j_{n'}}{j_{p'}}{I}
\rho^{pp}_{I^\pi}(pp'nn';I^\pi).
\label{40}\end{eqnarray}

The reduced single-particle pn form factors for the one-body operators
defined in \rf{26} are \cite{Ros54,Def66}
\begin{eqnarray}
\Bra{p}{\sf Y}^\kappa_{\lambda J}(k)\Ket{n}&=&(4\pi)^{-{1 \over 2}}
W_{J0J}(pn)R^\kappa_\lambda(pn;k),
\nonumber\\
\Bra{p}{\sf S}_{\lambda LJ}^\kappa(k)\Ket{n}&=&(4\pi)^{-{1\over 2}}
W_{L1J}(pn)R^\kappa_\lambda(pn;k),
\label{41}\\
\Bra{p}{\sf P}_{LJ}(k)\Ket{n}&=&(4\pi)^{-{1 \over 2}}
\left[W_{LJ}^{(-)}(pn) R_L^{(-)}(pn;k)
+W_{LJ}^{(+)}(pn)R_L^{(+)}(pn;k)\right],
\nonumber\end{eqnarray}
with the angular parts\footnote{We use here the angular momentum coupling 
$\ket{({1 \over 2},l)j}$.}
\begin{eqnarray}
W_{LSJ}(pn)&=&\sqrt{2}\hat{S} \hat{J}\hat{L}\hat{l}_n\hat{j}_n\hat{j}_p
(l_nL|l_p) \ninj{l_p}{{1 \over 2}}{j_p}{L}{S}{J}{l_n}{{1 \over 2}}{j_n},
\nonumber\\
W_{LJ}^{(\pm)}(pn)&=&\mp i(-1)^{l_p+j_n+J+{1 \over 2}} \hat{J}\hat{L}
\hat{l}_p
\hat{j}_p\hat{j}_n (l_n+{\ss{1 \over 2}}\mp
{\ss{1 \over 2}})^{{1 \over 2}}(l_pL|l_n\mp 1)\nonumber\\
&{\times}&\sixj{l_p}{j_p}{{1 \over 2}}{j_n}{l_n}{J}
\sixj{L}{J}{1}{l_n} {l_n\mp 1}{l_p},
\label{42}\end{eqnarray}
and the radial parts
\begin{eqnarray}
R^\kappa_L(pn;k) &\equiv&R^\kappa_L(l_p,n_p,l_n,n_n;k)=
\int_0^\infty u_{n_p,l_p}(r)u_{n_n,l_n}(r)j_L(kr)r^{2+\kappa} dr,
\nonumber\\
R_L^{(\pm)}(pn;k)&=&\int_0^\infty u_{n_p,l_p}(r)\left(\frac{d}{dr}\pm
\frac{2l_n+1\pm 1}{2r}\right)u_{n_n,l_n}(r)j_L(qr)r^2 dr.
\label{43}\end{eqnarray}

To carry out the numerical calculation of eqs. (27)~-~(33) it is convenient
to group separately the angular and the radial parts. For instance,
$M_{GT}$ can be cast in the form
\begin{eqnarray}
M_{GT}&=&-\sum_{LJ^\pi_{\a}}(-)^L\sum_{pp'nn'}
\rho^{ph}(pnp'n';J_\a^\pi)W_{L1J}(pn) W_{L1J}(p'n')
{\cal R}^0_{LL}(pnp'n';{\omega}_{J^\pi_{\a}}),
\label{44}\end{eqnarray}
where the two-body radial integrals are defined as
\begin{equation}
{\cal R}^\kappa_{LL'}(pnp'n';{\omega}_{J^\pi_{\a}})={\sf R}\int dk k^{2+\kappa}
 v(k;{\omega}_{J^\pi_{\a}})R_L^0(pn;k) R_{L'}^0(p'n';k).
\label{45}\end{equation}

One manner to include the effects of the finite nucleon size (FNS) and the
two-nucleon short-range correlations (SRC) on the $\beta\beta_{0\nu}$
moments has been explicated in ref. \cite{Krm92}, with the result:
\begin {eqnarray}
v(k,\omega_{{\sss {N}}})&\rightarrow&
v(k,\omega_{{\sss {N}}})
\left(\frac{\Lambda^2}{\Lambda^2+k^2}\right)^4
-\frac{1}{\pi kk_{c}}\ln\left| {\frac{k+k_{c}}{k-k_{c}}}\right|
\nonumber\\
&+&\frac{1}{2\pi kk_{c}}\left[\sum_{n=1}^{3}\frac{1}{n}
\left(x_{-}^{n}-x_{+}^{n}\right)+\ln\left(\frac{x_{-}}{x_{+}}\right)\right],
\label{46} \end{eqnarray}
where $\Lambda=850$~MeV is the cutoff for the dipole form factor in the FNS
correlations,
\begin{equation}
x_{\pm}=\frac{\Lambda^{2}}{\Lambda^{2}+(k\pm k_{c})^{2}},
\label{47}\end{equation}
and $k_c=3.93$ fm$^{-1}$ is roughly the Compton wavelength of the
$\omega$-meson in the SRC correlations.

The integration on the neutrino momentum $k$ is simplified when the
harmonic oscillator radial wave functions are employed. Then
the following relations among the one-body radial integrals are valid:
\begin{eqnarray}
R^1_L(pn;k) &=&(2\nu)^{-{1 \over 2}}
\left\{(2l_n+2n_n+3)^{{1 \over 2}}R^0_L(k;l_p,n_p,l_n+1,n_n)\right.
\nonumber\\
&-&\left.(2n_n)^{{1 \over 2}}R^0_L(k;l_p,n_p,l_n+1,n_n-1)\right\},
\nonumber\\
R_L^{(\pm)}(pn;k)&=&\pm\left(\frac{\nu}{2}\right)^{{1 \over 2}}
\left\{(2l_n+2n_n+2\mp 1)^{{1 \over 2}}R^0_L(k;l_p,n_p,l_n\mp 1,n_n)
\right.\nonumber\\
&+&\left.(2n_n+1\pm 1)^{{1 \over 2}}R^0_L(k;l_p,n_p,l_n\mp 1,n_n\pm 1)
\right\},
\label{48}\end{eqnarray}
where $\nu=M\omega/\hbar$ is the oscillator parameter, and the
$k$-integration in the matrix elements \rf{27}~-~\rf{33} only involves the
radial integrals \rf{45}.
Their  explicit forms in this case are shown in the appendix D.

The densities $\rho^{ph}(pnp'n';J^\pi_\a)$ and
$\rho^{ph}_{cl}(pnp'n';J^\pi)$ are supplied by the nuclear
structure calculations. As an example, we
discuss below the $\beta\beta$ decay $^{48}Ca\rightarrow  ^{48}\!Ti$.

We first consider the case when the intermediate nucleus $^{48}Sc$  and
the final nucleus $^{48}Ti$ are described, respectively, as one-particle
one-hole and
two-particle two-hole excitations on $^{48}Ca$, {\em i.e., }
\begin{eqnarray}
\ket{J^\pi_{\a}M}&=&\sum_{pn} \ov{pn}{J^\pi_{\a}}
\left(a^{{\dagger}}_p a_{\bar{n}}\right)_{J^\pi M}\ket{0^+_i},
\nonumber\\
\ket{0^+_f}&=&
\sum_{p\geq p'n\geq n' I^{\pi}}
N(pp')N(nn')
\ov{pp'nn';I^{\pi}}{0^+_f}\left[(a^{{\dagger}}_pa^{{\dagger}}_{p'}
)_{I^{\pi}}(a_{\bar{n}} a_{\bar{n}'})_{I^{\pi}}\right]_0
\ket{0^+_i},
\label{49}\end{eqnarray}
with $N(pp')=(1+\delta_{pp'})^{-\frac{1}{2}}$.
One gets from \rf{37}
\begin{eqnarray}
{\rho}^{ph}(p_1n_1p_2n_2;J_{\a}^\pi)
&=&\ov{J^\pi_{\a}}{p_2n_2}
\sum_{p\geq p'n\geq n'} \sum_{p_3n_3 I^{\pi}}\hat{I}
\ov{p_3n_3}{J^\pi_{\a}} \ov{0^+_f}{pp'nn';I^{\pi}}
(-)^{n_1+p_3+J+I}\nonumber\\
&{\times}&\sixj{p_1}{n_1}{J}{n_3}{p_3}{I}\hat{P}_{I}(pp')\hat{P}_{I}(nn')
\delta_{pp_1}\delta_{nn_1} \delta_{p'p_3}\delta_{n'n_3},
\label{50}\end{eqnarray}
\newpage
and
\begin{eqnarray}
\rho^{ph}_{cl}(p_1n_1p_2n_2;J^\pi) &=& \sum_{p\geq p'n\geq n' I^{\pi}}
\hat{I}\ov{0^+_f}{pp'nn';I^{\pi}}
(-)^{n_1+p_2+J+I}\nonumber\\
&{\times}&\sixj{p_1}{n_1}{J}{n_2}{p_2}{I}\hat{P}_{I}(pp')\hat{P}_{I}(nn')
\delta_{pp_1}\delta_{nn_1} \delta_{p'p_2}\delta_{n'n_2},
\label{51}\end{eqnarray}
where
\begin{equation}
\hat{P}_J(pp')=N(pp')\left[1-(-)^{J+p+p'}(p\leftrightarrow p')\right].
\label{52}\end{equation}
Within the CA one can use the closure relation
\begin{equation}
\sum_{\a}\ov{p_3n_3}{J^\pi_{\a}} \ov{J^\pi_{\a}}{p_2n_2}
=\delta_{p_2p_3}\delta_{n_2n_3},
\label{53}\end{equation}
which leads from \rf{50} to \rf{51}.

On the other hand, within the QRPA formulation, and after solving the BCS
equations for the intermediate nucleus $^{48}Sc$ \cite{Krm96}, the two-body
density matrix becomes
\begin{eqnarray}
{\rho}^{ph}(pnp'n';J_{\a}^\pi)=
\left[u_nv_pX_{J^\pi_{\a}}(pn)+ u_pv_nY_{J_{\a}^\pi}(pn)\right]
\left[ u_{p'}v_{n'}X_{J^\pi_{\a}}(p'n')+u_{n'}v_{p'}Y_{J^\pi_{\a}}(p'n')
\right],
\label{54}\end{eqnarray}
where all the notation has the standard meaning \cite{Krm94,Krm96}.
One should bear in mind that when the QRPA is used, the energies
${\omega}_{J^\pi_{\a}}$ that appear in the matrix radial integrals are the
solutions of the RPA problem and {\em not} the excitation energies
of the intermediate nucleus relative to the initial nucleus.

In particular, in the single mode model \cite{Krm92},
where there is only one intermediate state for each $J^\pi$
(and which seems to be a reasonable first order approximation for the
$\beta\beta$ decays of $^{48}Ca$ and $^{100}Mo$ nuclei \cite{Krm94a,Eji96}),
\begin{eqnarray}
\rho^{ph}(pnpn;J^\pi)
&=&u_pv_nu_nv_p\left(\frac{{\omega}^0} {{\omega}_{J^\pi}}\right)
\left(1+\frac{G(J^\pi)}{{\omega}_0}\right),
\label{55}\end{eqnarray}
where $G(J^\pi)=G(pnpn;J^\pi)$, 
${\omega}^0=-\left[G(pppp;0^+)+ G(nnnn;0^+)\right]/4$,
and $=G(jj'jj';J^\pi)$ are the particle-particle matrix elements.
The intermediate states for $^{48}Sc$ are:\\
$[0f_{7/2}(p)0f_{7/2}(n)]_{J^+}$,
and the values of the ratios $G(J^+)/{\omega}_0$ for the $\delta$ force
can be found in table 1 of ref. \cite{Krm94a}. 

\begin{center}
\section{Summarizing Discussion }
\end{center}

A straightforward derivation of the $\beta\beta_{0\nu}$ decay rate, based on
the Fourier-Bessel expansion of the transition amplitude, and the posterior
application of the Racah algebra, has been performed without invoking the
closure approximation.
If necessary, this approximation can be implemented, however, at any step
of the calculation.
It has been used for deriving the $\beta\beta_{0\nu}$ formulas in refs.
\cite{Hax84,Doi85,Tom86}, but not in refs. \cite{Ver90,Suh91}.

To evaluate the nuclear matrix elements exhibited in eqs. (27)~-~(33) we only
have to perform summations on the angular momenta and the intermediate
virtual states. The successive terms rapidly decrease, because the radial
integrals \rf{45} steadily diminish in magnitudes when the multipolarities
$L$ and $L'$ are augmented \cite{Krm92,Krm94}.
The formulas become particularly simple when the harmonic oscillator basis
is used. Then the Horie and Sasaki method \cite{Hor61} can be exploited
for the evaluation of the radial form factors \rf{45},
and the equations displayed in the appendix D can be used.

The present formalism is especially suitable for the nuclear structure
in which the summation on the intermediate states is unavoidable,
such as the QRPA. The closure approximation  then just connotes that
the variation of the energy denominators with nuclear excitation
is not considered. Evidently this does not lead to a major simplification in
the numerical calculation.
For example, because of ${\rho}^{ph}(pnp'n';J_{\a}^\pi)$ given by \rf{54},
the summation in \rf{44} on different states ${\a}$  with the same $J^\pi$
persists, although we do the replacement
${\omega}_{J^\pi_{\a}}\rightarrow <{\omega}_{J^\pi_{\a}}> ={\omega}_{J^\pi}$.

Contrarily, the use of the closure approximation
is mandatory, and can be implemented easily as described in the last section,
when the study is done in the shell-model framework, i.e., when one
possesses information only on ${\rho}^{ph}_{cl}$, or equivalently on the $0^+$
nuclear wave functions for the initial and final states.
In this case the matrix element \rf{44} reads
\[
M_{GT}=-\sum_{LJ^\pi}(-)^L\sum_{pp'nn'}
\rho^{ph}_{cl}(pnp'n';J^\pi)W_{L1J}(pn) W_{L1J}(p'n')
{\cal R}^0_{LL}(pnp'n';{\omega}_{J^\pi}).\]

The main difference between the formalism presented here and those published
so far \cite{Hax84,Doi85,Tom86,Ver90,Suh91} is its simplicity.
As such it is more suitable for the numerical calculations.
Let us underscore a few points in this regard:

1) While in the neutrino potential formalisms \cite{Hax84,Doi85,Tom86} one deals
with two-body matrix elements, which lead to rather complicated analytic
expressions for the
$\beta\beta_{0\nu}$ moments,
we only have to handle the well known one-body operators
\rf{26}.
It could be illustrative to compare our result \rf{31} for the matrix element
$M_{\ss R}$ with eqs. (3.65) to (3.68) in the Tomoda's report \cite{Tom86}.

2) At variance with the formalism developed by Vergados {\it et al.},
\cite{Ver90}, the results shown here are not limited to the employment of
harmonic oscillator one-particle wave functions. Besides we totally avoid
the usage of the Moshinsky-Brody transformation brackets, which now and
then could be cumbersome.

3) There are as well several substantial differences with the works of
Suhonen {\it et al.} \cite{Suh91}, where the Fourier-Bessel
expansion has also been used. First, they obtain different and more complex results
for $M_{F}$ and $M_{GT}$. Second, they do not exhibit the explicit structure
of for the remaining matrix elements, given here by eqs.  (29)~-~(33),
but only show their general layout. Yet, this layout cannot be used for any
practical purpose.
Third, instead of dealing with the plain nuclear shell model,
they operate in a
relativistic quark confinement model.
Fourth, their formulation is limited to the QRPA approximation
as well as to the harmonic oscillator basis.

In summary, we believe that the present formalism simplifies
the nuclear structure evaluation of the $\beta\beta_{0\nu}$ matrix elements
to a large extent. The formulation is applicable as well to
matrix elements that appear in some supersymmetric contributions
\cite{Hir95}.

\pagebreak
\renewcommand{\theequation}{\thesection.\arabic{equation}}
\appendix
\setcounter{equation}{0}
\begin{center}
\subsection*{Appendix A:
Matrix elements ${\sf M}_{X}({\bf k},{{\ss N}})$}
\end{center}

\setcounter{section}{1}

After integrating on $d{\bf x}$ and $d{\bf y}$, as indicated
in eq. \rf{9}, the matrix elements \rf{11} and \rf{12} read
\begin{eqnarray}
{\sf M}_{F}({\bf k},{{\ss N}})&=&g_{{\sss V}}^2{\bra{{\ss F}}}\sum_n\tau_n^+
e^{i{\bf k}\cdot{\bf r}_n}\ket{{\ss N}}\bra{{\ss N}}\sum_m\tau_m^+
e^{-i{\bf k}\cdot{\bf r}_m}\ket{{\ss I}},
\label{A.1}\end{eqnarray}
\begin{eqnarray}
{\sf M}_{GT}({\bf k},{{\ss N}})&=&g_{{\sss A}}^2  {\bra{{\ss F}}}\sum_n\tau_n^+\mbs_n
e^{i{\bf k}\cdot{\bf r}_n}\ket{{\ss N}}\cdot\bra{{\ss N}}\sum_m\tau_m^+\mbs_m
e^{-i{\bf k}\cdot{\bf r}_m}\ket{{\ss I}},
\label{A.2}\end{eqnarray}
\begin{eqnarray}
{\sf M}'_{F}({\bf k},{{\ss N}})&=&-2ig_{{\sss V}}^2
{\bra{{\ss F}}}\sum_n\tau_n^+ {\bf k}\cdot{\bf r}_ne^{i{\bf k}
\cdot{\bf r}_n}\ket{{\ss N}}\bra{{\ss N}}\sum_m \tau_m^+e^{-i{\bf k}
\cdot{\bf r}_m}\ket{{\ss I}},
\label{A.3}\end{eqnarray}
\begin{eqnarray}
{\sf M}'_{GT}({\bf k},{{\ss N}})&=&-2ig_{{\sss A}}^2
{\bra{{\ss F}}}\sum_n\tau_n^+ {\bf k}\cdot{\bf r}_n\mbs_ne^{i{\bf k}
\cdot{\bf r}_n}\ket{{\ss N}}\cdot\bra{{\ss N}}\sum_m \tau_m^+\mbs_m
e^{-i{\bf k}\cdot{\bf r}_m}\ket{{\ss I}},
\label{A.4}\end{eqnarray}
\begin{eqnarray}
{\sf M}_{{\sss R}}({\bf k},{{\ss N}})&=&
-i\frac{{\sf R}g_{{\sss A}}g_{{\sss V}}}{2M_{\ss N}}
{\bf k}\cdot\left\{ {\bra{{\ss F}}}\sum_n\tau_n^+\mbs_n
e^{i{\bf k}\cdot{\bf r}_n}\ket{{\ss N}}{\times}\right.
\nonumber\\
&&\left.\bra{{\ss N}}\sum_m\tau_m^+
\left[{\bf p}_me^{-i{\bf k}\cdot{\bf r}_m}+e^{-i{\bf k}\cdot
{\bf r}_m}{\bf p}_m
+f_{\sss W}{\mbox{\boldmath$\nabla$}}{\times}\mbs_m
e^{-i{\bf k}\cdot{\bf r}_m}\right]\ket{{{\ss I}}}\right\},\nonumber\\
\label{A.5}\end{eqnarray}
\begin{eqnarray}
{\sf M}_{T}({\bf k},{{\ss N}})&=&\frac{2\sqrt{5}}{\sqrt{3}}ig_{{\sss A}}^2  
{\bra{{\ss F}}}\sum_n\tau_n^+
[\mbs_n\otimes({\bf k}\otimes{\bf r}_n)^{(2)}]^{(1)}e^{i{\bf k}\cdot{\bf r}_n}
\ket{{\ss N}}\cdot\bra{{\ss N}}\sum_m\tau_m^+\mbs_m
e^{-i{\bf k}\cdot{\bf r}_m}\ket{{\ss I}},\nonumber\\
\label{A.6}\end{eqnarray}
\begin{eqnarray}
{\sf M}_{P}({\bf k},{{\ss N}})&=&-\sqrt{2}ig_{{\sss A}}g_{{\sss V}}  
\left\{\sqrt{3}{\bra{{\ss F}}}\sum_n\tau_n^+
[\mbs_n\otimes({\bf k}\otimes{\bf r}_n)^{(1)}]^{(0)}e^{i{\bf k}\cdot{\bf r}_n}
\ket{{\ss N}}\bra{{\ss N}}\sum_m\tau_m^+e^{-i{\bf k}\cdot{\bf r}_m}\ket{{\ss I}}
\right.\nonumber\\
&-&\left.{\bra{{\ss F}}}\sum_n\tau_n^+\mbs_ne^{i{\bf k}\cdot{\bf r}_n}
\ket{{\ss N}}\cdot\bra{{\ss N}}\sum_m\tau_m^+({\bf k}\otimes{\bf r}_m)^{(1)}
e^{-i{\bf k}\cdot{\bf r}_m}\ket{{\ss I}}\right\}.
\label{A.7}\end{eqnarray}
\setcounter{equation}{0}
\begin{center}
\subsection*{ Appendix B: Electron matrix elements and phase-space factors}
\end{center}
\setcounter{section}{2}
The leptonic factors in eq. \rf{18} are:
\begin{eqnarray}
L_1(\epsilon_1,\epsilon_2)&=&(-1)^{1/2-s'_2}\chi_{s'_1}^{\dagger}
[g_{-1}(\epsilon_1)-
f_{1}(\epsilon_1)\mbs\cdot\hat{{\bf p}}_1][f_{1}(\epsilon_2)
\mbs\cdot\hat{{\bf p}}_2+
g_{-1}(\epsilon_2)]\chi_{-s'_2},\nonumber\\
L_2(\epsilon_1,\epsilon_2)&=&(\epsilon_1-\epsilon_2)(-1)^{1/2-s'_2}
\chi_{s'_1}^{\dagger}[g_{-1}(\epsilon_1)
f_{1}(\epsilon_2)\mbs\cdot\hat{{\bf p}}_2+
f_{1}(\epsilon_1)g_{-1}(\epsilon_2)\mbs\cdot\hat{{\bf p}}_1]\chi_{-s'_2}.
\nonumber\\
L_3(\epsilon_1,\epsilon_2)&=&{1\over{\sf R}}(-1)^{1/2+s'_2}\chi_{s'_1}^{\dagger}
[g_{-1}(\epsilon_1)g_{-1}(\epsilon_2)
+f_{1}(\epsilon_1)f_{1}(\epsilon_2)\mbs\cdot\hat{{\bf p}}_1
\mbs\cdot\hat{{\bf p}}_2]\chi_{-s'_2}\label{B.1}\\
L_4(\epsilon_1,\epsilon_2)&=&\frac{i}{2{\sf R}}(-1)^{1/2+s'_2}
\chi_{s'_1}^{\dagger}
\left\{[f_{1}(\epsilon_1)f_{-1}(\epsilon_2)+g_{1}(\epsilon_1)
g_{-1}(\epsilon_2)]\mbs\cdot\hat{{\bf p}}_1
\right.\nonumber\\
&-&\left.[f_{-1}(\epsilon_1)f_{1}(\epsilon_2)+g_{-1}(\epsilon_1)
g_{1}(\epsilon_2)]\mbs\cdot\hat{{\bf p}}_2
\right\}\chi_{-s'_2},\nonumber\\
L_5(\epsilon_1,\epsilon_2)&=&\frac{i}{2{\sf R}}(-1)^{1/2+s'_2}
\chi_{s'_1}^{\dagger}
\left\{[g_{-1}(\epsilon_1)f_{-1}(\epsilon_2) +f_{-1}(\epsilon_1)
g_{-1}(\epsilon_2)]\right.\nonumber\\
&-&\left.[g_{1}(\epsilon_1)f_{1}(\epsilon_2) +f_{1}(\epsilon_1)
g_{1}(\epsilon_2)]
\mbs\cdot\hat{{\bf p}}_1 \mbs\cdot\hat{{\bf p}}_2\right\}\chi_{-s'_2},
\nonumber\end{eqnarray}
where all the notation has the usual meaning
\cite{Doi85}.

The electron phase-space factors
${\cal F}_k(T_0)$ that appear in eq. \rf{24} are:
\begin{eqnarray}
{\cal F}_1(T_0)&=&T_0(30+60T_0+40T_0^2+10T_0^3+T_0^4)/30,\nonumber\\
{\cal F}_2(T_0)&=&T_0^4(70+77T_0+14T_0^2+T_0^3)/420,\nonumber\\
{\cal F}_3(T_0)&=&T_0^3(10+10T_0+T_0^2)/30,\nonumber\\
{\cal F}_4(T_0)&=&T_0^2(30+35T_0+10T_0^2+T_0^3)/135,\nonumber\\
{\cal F}_5(T_0)&=&T_0[60T_0+80T_0^2+30T_0^3+3T_0^4+\xi
(60+90T_0+40T_0^2+5T_0^3)]/45,
\label{B.2}\\
{\cal F}_6(T_0)&=&2T_0(12+18T_0+8T_0^2+T_0^3)/(3{\sf R}),\nonumber\\
{\cal F}_7(T_0)&=&4T_0[60T_0+100T_0^2+55T_0^3+12T_0^4+T_0^5+
\xi(60+90T_0+45T_0^2+10T_0^3+T_0^4)]/(45 {\sf R}),\nonumber\\
{\cal F}_8(T_0)&=&T_0[100T_0^2+150T_0^3+73T_0^4+14T_0^5+T_0^6
+2\xi(60T_0+100T_0^2+55T_0^3+12T_0^4+T_0^5)\nonumber\\
&+&
\xi^2(60+90T_0+45T_0^2+10T_0^3+T_0^4)]/135,\nonumber\\
{\cal F}_9(T_0)&=&4T_0(60+90T_0+45T_0^2+10T_0^3+T_0^4)/(15{\sf R}^2),
\nonumber\end{eqnarray}
with
\begin{equation}
\xi=\frac{3\a Z}{{\sf R}}.
\label{B.3}\end{equation}

\pagebreak
\setcounter{equation}{0}
\begin{center}
\subsection*{ Appendix C: Derivation of the final formulas for the nuclear
moments}
\end{center}
\setcounter{section}{3}

Below we give the details on the derivation of the last term in \rf{31}.
First we rewrite \rf{A.5} as
\begin{eqnarray}
{\sf M}_{{\sss R}}({\bf k},{{\ss N}})&=&
-i\frac{{\sf R}g_{{\sss A}}g_{{\sss V}}} {2M_{\ss N}}
{\bf k}\cdot{\bra{\ss F}}\sum_n\tau_n^+\mbs_n e^{i{\bf k}\cdot
{\bf r}_n}\ket{{\ss N}}{\times}
\nonumber\\
&&\bra{{\ss N}}\sum_m\tau_m^+ e^{-i{\bf k}\cdot{\bf r}_m}
\left[2{\bf p}_m-{\bf k} +if_{\sss W}\mbs_m{\times}{\bf k}\right]
\ket{{\ss I}},
\label{C.1} \end{eqnarray}
and then we express the vector product, involving the nucleon momentum
term in \rf{C.1}, in spherical coordinates
\begin{eqnarray}
{\sf M}_{{\sss R}}^{({\bf p})}({\bf k},{{\ss N}})&=&
\frac{{\sf R}g_{{\sss A}}g_{{\sss V}}} {M_{\sss N}}
\sqrt{2}\sum_{\nu\nu'\mu}(-1)^{\mu}
(1\nu1\nu'|1\mu)k^\mu
\nonumber\\
&&{\bra{{\ss F}}}\sum_n\tau_n^+\sigma^{-\nu}_n
e^{i{\bf k}\cdot{\bf r}_n}
\ket{{\ss N}} \bra{{\ss N}}\sum_m\tau_m^+ e^{-i{\bf k}\cdot{\bf r}_m}
p^{-\nu'}_m  \ket{{\ss I}}.
\label{C.2} \end{eqnarray}
After performing the
multipole expansion \rf{25} and handling some straightforward
Racah algebra, we obtain
\begin{eqnarray}
&&{\sf M}_{{\sss R}}^{({\bf p})}({\bf k},{{\ss N}})=
\frac{{\sf R}g_{{\sss A}}g_{{\sss V}}} {M_{\ss N}}\sqrt{6} (4\pi)^2k
\sum_{LL'JJ'M'{\rm M}{\rm M}'\kappa\rho\nu}i^{L-L'}(-1)^{{\rm M}'}\hat{L}
\sixj{L}{J'}{1}{1}{1}{\kappa}(L010|\kappa 0)
\nonumber\\
&&(1,-\nu\kappa\rho|J',-{\rm M}')(1,-\nu L'M'|J{\rm M})
Y_{\kappa\rho}({\hat{{\bf k}}})Y_{L'M'}^*({\hat{{\bf k}}})
\bra{{\ss F}}{\sf S}_{LLJ'{\rm M}'}^0(k)\ket{{\ss N}}
\bra{{\ss N}}{\sf P}_{L'J{\rm M}}(k)\ket{{\ss I}},
\nonumber\\\label{C.3} \end{eqnarray}
where the tensor operators ${\sf S}_{LLJ'}^0(k)$
and ${\sf P}_{L'J}(k)$ are defined in \rf{26}.
Finally, the angular integration allows us to perform the summations on the
angular momentum projections and obtain
\begin{eqnarray}
\int d\Omega_{{\bf k}}{\sf M}_{{\sss R}}^{({\bf p})}({\bf k},{{\ss N}})&=&
\frac{{\sf R}g_{{\sss A}}g_{{\sss V}}} {M_{\ss N}}
\sqrt{6}(4\pi)^2k\sum_{LL'J}i^{L-L'}\hat{L}
\sixj{L}{J}{1}{1}{1}{L'} (L010|L'0)\nonumber\\
&&\bra{{\ss F}}{\sf S}_{LLJ}^0(k)\ket{{\ss N}}\cdot
\bra{{\ss N}}{\sf P}_{L'J}(k)\ket{{\ss I}}.
\label{C.4}\end{eqnarray}
This result, together with the eq. \rf{14}, yields the last term in
eq. \rf{31}.

\pagebreak

\setcounter{equation}{0}
\begin{center}
\subsection*{Appendix D:
Radial form factors for the harmonic oscillator wave functions}
\end{center}

\setcounter{section}{4}

Following the Horie and Sasaki
method \cite{Hor61} the radial integral \rf{45} can be expressed as:
\begin{equation}
{\cal R}^\kappa_{LL'}(pnp'n'; {\omega}_{J^\pi_{\a}})
=[M(p,n)M(p',n')]^{-1/2}\sum_{mm'} a_m(p,n)a_{m'}(p',n')
f^\kappa_{LL'}(m,m'; {\omega}_{J^\pi_{\a}}),
\label{D.1}\end{equation}
where
\begin{eqnarray}
M(n_pl_p,n_nl_n)&=&2^{n_p+n_n}n_p!n_n!(2l_p+2n_p+1)!!(2l_n+2n_n+1)!!,
\label{D.2}\end{eqnarray}
\begin{eqnarray}
a_{l_p+l_n+2s}(n_pl_p,n_nl_n)&=&\sum_{k+k'=s}\left(\begin{array}{c}n_p\\
k\end{array}\right) \left(\begin{array}{c}n_n\\
k'\end{array}\right)\frac{(2l_p+2n_p+1)!!(2l_n+2n_n+1)!!}
{(2l_p+2k+1)!!(2l_n+2k'+1)!!},
\nonumber\\\label{D.3}\end{eqnarray}
\begin{eqnarray}
f^\kappa_{LL'}(m,m'; {\omega}_{J^\pi_{\a}})
&=&\sum_\mu a_{2\mu}\left(\frac{m-L}{2} L,
\frac{m'-L'}{2} L'\right){\cal{J}}^\kappa_\mu({\omega}_{J^\pi_{\a}}),
\label{D.4}\end{eqnarray}
and
\begin{equation}
{\cal{J}}^\kappa_\mu( {\omega}_{J^\pi_{\a}})
=(2\nu)^{-\mu}{\sf R}\int_0^\infty dk k^{2\mu+2+\kappa}
e^{-k^2/2\nu}v(k;{\omega}_{J^\pi_{\a}}).
\label{D.5}\end{equation}

\end{document}